# MOVPE-grown Si-doped $\beta$-(Al$_{0.26}$Ga$_{0.74}$)$_2$O$_3$ thin films and heterostructures

Praneeth Ranga[1]*, Ashwin Rishinaramangalam[2], Joel Varley[3], Arkka Bhattacharyya[1], Daniel Feezell[2], Sriram Krishnamoorthy[1]*

[1] *Department of Electrical and Computer Engineering, The University of Utah, Salt Lake City, UT, 84112, USA*

[2] *Center for High Technology Materials, University of New Mexico, Albuquerque, NM 87106, USA*

[3] *Lawrence Livermore National Laboratory, Livermore, CA, 94550, United States of America*

E-mail: praneeth.ranga@utah.edu; sriram.krishnamoorthy@utah.edu

We report on n-type degenerate doping in MOVPE grown $\beta$-(Al$_{0.26}$Ga$_{0.74}$)$_2$O$_3$ epitaxial thin films and modulation doping in $\beta$-(Al$_{0.26}$Ga$_{0.74}$)$_2$O$_3$/$\beta$-Ga$_2$O$_3$ heterostructure. Alloy composition is confirmed using HRXRD measurements. Carrier concentration in the thin films is proportional to the silane molar flow. Room temperature hall measurements showed a high carrier concentration of $6\times10^{18}$-$7.3\times10^{19}$ cm$^{-3}$ with a corresponding electron mobility of 53-27 cm$^2$/V.s in uniformly-doped $\beta$-(Al$_{0.26}$Ga$_{0.74}$)$_2$O$_3$ layers. Modulation doping is used to realize a total electron sheet charge of $2.3\times10^{12}$ cm$^{-2}$ in a $\beta$-(Al$_{0.26}$Ga$_{0.74}$)$_2$O$_3$/$\beta$-Ga$_2$O$_3$ heterostructure using a uniformly-doped $\beta$-(Al$_{0.26}$Ga$_{0.74}$)$_2$O$_3$ barrier layer and a thin spacer layer.

β-$Ga_2O_3$ is an emerging ultra-wide band gap semiconductor with potential applications in high power electronics[1] and deep UV photodetectors.[2] Considering a room temperature mobility of 200-300 $cm^2$/V.s and predicted breakdown field of 6-8 MV/cm, β-$Ga_2O_3$ has much higher predicted BFOM (Baliga figure of merit) than other compound semiconductors such as GaN and SiC. A critical breakdown field higher than GaN has already been experimentally demonstrated.[3] The high BFOM, availability of high quality single-crystal bulk substrates[4], demonstration of shallow n-type substitutional donors and ability to achieve a wide range of conductivity make it an attractive material for high power devices.[1] Breakdown voltages exceeding 2kV have already been demonstrated in both lateral and vertical epitaxial devices, indicating high potential for β-$Ga_2O_3$ based power devices.[5,6] In addition to high power applications, β-$Ga_2O_3$ has a high JFOM (Johnson figure of merit), making it a potential material for RF power amplifiers.[7,8]

β-$(Al_xGa_{1-x})_2O_3$ is a monoclinic ternary alloy with a band gap in the range of 4.8 – 6.2 eV for $Al_2O_3$ mole fraction of 0 to 71%.[9] The BFOM of ultra-wide band gap β-$(Al_xGa_{1-x})_2O_3$ is expected to be even higher than that of $Ga_2O_3$. This makes β-$(Al_xGa_{1-x})_2O_3$ a potential candidate for high power electronic devices. Epitaxial thin films with composition up to x = 0.2 have already been demonstrated using MBE.[10] A preliminary report on β-$(Al_xGa_{1-x})_2O_3$ MESFETs already shows higher critical breakdown field compared to β-$Ga_2O_3$.[11] However, they suffer from low mobility which leads to lower on current in the device. It is therefore critical to study doping and improve mobility in β-$(Al_xGa_{1-x})_2O_3$ in order to understand electronic transport in β-$(Al_xGa_{1-x})_2O_3$ alloys.

Study of doping in β-$(Al_xGa_{1-x})_2O_3$ is also important, as β-$(Al_xGa_{1-x})_2O_3$ is used as a barrier layer for β-$(Al_xGa_{1-x})_2O_3$/β-$Ga_2O_3$ HEMT (High Electron Mobility transistors). β-$Ga_2O_3$ suffers severe polar optical phonon(POP) scattering, limiting its maximum theoretical mobility to 200 – 300 cm$^2$/V.s at room temperature.[12] Theoretical studies indicate that confining a high density of carriers at β-(Al$_x$Ga$_{1-x}$)$_2$O$_3$/β-Ga$_2$O$_3$ interface can lead to mobilities much higher than that of bulk β-Ga$_2$O$_3$. The enhanced 2DEG sheet charge is predicted to screen out certain phonon modes, resulting in reduced scattering in the 2DEG channel.[13] Recently, modulation doped β-(Al$_x$Ga$_{1-x}$)$_2$O$_3$/β-Ga$_2$O$_3$ HEMT's with a sheet charge of 2x10$^{12}$ cm$^{-2}$ and Hall mobility of 180 cm$^2$/V.s have been demonstrated.[14] However, the maximum reported 2DEG sheet charge without parallel channel is nearly 2x10$^{12}$ cm$^{-2}$ and, is being limited by the maximum Al mole fraction attainable in the barrier layer. With a large conduction band offset, a larger 2DEG density can be confined in the channel without the formation of a parasitic parallel channel in the alloy barrier. Ab-initio calculations show β-(Al$_x$Ga$_{1-x}$)$_2$O$_3$ up to x = 0.71 is thermodynamically stable and ΔE$_c$ as high as 1.5 eV can be achieved using a β-(Al$_{0.8}$Ga$_{0.2}$)$_2$O$_3$/β-Ga$_2$O$_3$ heterojunction, resulting in a larger design space for heterostructure materials and devices.[9] So far, modulation doping has been studied only using MBE,[14–16] which limits the maximum attainable Al mole fraction due to issues with gallium incorporation at higher temperatures required for growth of β-(Al$_x$Ga$_{1-x}$)$_2$O$_3$ (>800°C), necessitating other approaches such as MOCATAXY(Metal-oxide catalyzed epitaxy).[10] MOVPE growth has advantages such as larger growth temperature window and precise precursor flow control that makes it uniquely suitable for growth of high Al composition thin films and abrupt heterojunctions. MOVPE-grown epitaxial β-Ga$_2$O$_3$ films[17,18] have been already shown to have the highest experimentally measured mobility.[19] In this work, we study structural and electrical properties of heavily Si-doped β-(Al$_x$Ga$_{1-x}$)$_2$O$_3$ epitaxial thin films, which can lead to applications for high power and high frequency devices. We also demonstrate modulation doping via MOVPE growth using a doped β-(Al$_x$Ga$_{1-x}$)$_2$O$_3$/β-Ga$_2$O$_3$ abrupt heterojunction.

Growth of β-(Al$_x$Ga$_{1-x}$)$_2$O$_3$ epilayers is carried out using a far injection MOVPE reactor

(Agnitron Agilis) with TMAl, TEGa and $O_2$ as precursors and Ar as carrier gas. Prior to loading the substrates, the samples are solvent cleaned to remove any organic contaminants followed by a piranha etch. Growth is performed on diced 5x5 $mm^2$ Fe-doped (010) semi-insulating β-$Ga_2O_3$ and conducting Sn-doped bulk (010) β-$Ga_2O_3$ substrates (Novel Crystal Technology) at 810 °C with an oxygen flow of 500 sccm and a chamber pressure of 15 Torr. The Al composition in the thin films is controlled by fixing the [TMAl]/[TMAl+TMGa] molar ratio to 25%. N-type conductivity in the films is controlled by varying diluted [$SiH_4$]/[TMGa + TMAl] molar ratio. The growth rate of β-$(Al_xGa_{1-x})_2O_3$ is calibrated by growing a thick film on (010)-oriented Fe-doped bulk $Ga_2O_3$ substrate from Synpotics and measuring the cross section thickness in SEM. The growth rate of (010) β-$(Al_xGa_{1-x})_2O_3$ is found to be 3.8 nm/min, which correlates well with the thickness extracted from thin β-$(Al_xGa_{1-x})_2O_3$ films from high resolution X-ray diffraction measurements, discussed below.

We first examine electronic transport in heavily doped β-$(Al_xGa_{1-x})_2O_3$ epitaxial films. The epitaxial stack consists of a 150 nm thick undoped β-$Ga_2O_3$ buffer layer and a 16 nm linearly graded β-$(Al_xGa_{1-x})_2O_3$ from x = 0 to x, followed by a uniformly doped β-$(Al_xGa_{1-x})_2O_3$ layer (Fig. 1 inset). Direct growth of doped β-$(Al_xGa_{1-x})_2O_3$ on undoped β-$Ga_2O_3$ would result in formation of a modulation-doped channel in β-$Ga_2O_3$ and electron accumulation at the β-$(Al_xGa_{1-x})_2O_3$/β-$Ga_2O_3$ hetero interface. Therefore, in order to study doping in β-$(Al_xGa_{1-x})_2O_3$, we inserted a graded β-$(Al_xGa_{1-x})_2O_3$ layer,[20] so that the measured charge is contained in the doped β-$(Al_xGa_{1-x})_2O_3$ layer. This allows us to directly study electrical properties of β-$(Al_xGa_{1-x})_2O_3$ thin films.

Three samples of β-$(Al_xGa_{1-x})_2O_3$ (A, B and C) are grown using three different silane flows (Table 1), keeping the [TMAl]/[TMAl+TEGa] molar ratio the same and the other growth parameters (temperature, pressure, $O_2$ flow) the same. The Al composition of the thin films is estimated by comparing the peak separation between the (020) β-$(Al_xGa_{1-x})_2O_3$ and the bulk (020) β-$Ga_2O_3$ substrate peak from HRXRD measurements (Panalytical Empyrean), assuming a fully strained layer.[21] Samples A, B, and C show similar Al mole fraction (x ~ 0.26) (Fig. 1). The precursor molar flow ratio (25%) correlated very closely with the extracted alloy composition (26%) from HRXRD measurements, indicating no pre-reaction and complete incorporation of Al adatoms at this growth condition. Sample A showed thickness fringes in the HRXRD measurements corresponding to a thickness of 30 nm, which is close to the expected growth rate from the thick calibration sample. Samples B and C did not show any thickness fringe in the HRXRD scans, which is attributed to structural quality degradation due to higher silane flows/ doping concentration. The rms surface roughness of Sample A with 30 nm thick $\beta$-$(Al_{0.26}Ga_{0.74})_2O_3$ layer is 2.4 nm. The rms roughness of Samples B and C, with 40 nm thick $\beta$-$(Al_{0.26}Ga_{0.74})_2O_3$ increase to 4.8 nm and 5.7 nm, respectively (see supplementary data). We find that the surface roughness increases with the silane flow. It should be noted that uniformly doped $\beta$-$Ga_2O_3$ also show a rms surface roughness between 1-2 nm, indicating that further optimization of growth conditions is required to achieve atomically smooth surface morphology.

Ti (50 nm)/Au (50 nm) ohmic contacts are deposited on the four corners of the as-grown Si-doped $\beta$-$(Al_{0.26}Ga_{0.74})_2O_3$ thin films using DC sputtering. The as-deposited contacts are found to be ohmic at room temperature without the need for any additional annealing. Room temperature hall measurements (Ecopia HMS 7000) are used to characterize the sheet charge and carrier mobility as a function of temperature (80- 340K), after ascertaining the ohmic nature of the contacts. The results are summarized in Table 1. The room temperature carrier concentration is measured to be between $6 \times 10^{18}$ - $7 \times 10^{19}$ cm$^{-3}$ with a mobility of 53-27 cm$^2$/V.s. While the undoped $Ga_2O_3$ layer could contribute to the total measured hall charge and mobility, the contribution of the UID layer to measured hall conductivity is considered to be negligible.[22] The room temperature resistivity ($\rho$) for the samples is in the range of 0.019 - 0.0026 $\Omega$-cm, which is among the lowest reported values in n-type ultra wide band gap semiconductors of similar bandgap, such as $Al_{0.7}Ga_{0.3}N$.[20,23] Lower resistivity values (0.0013 Ω-cm) are reported in heavily doped (n – 1.7x10$^{20}$ cm$^{-3}$) pulsed laser deposited β-Ga$_2$O$_3$ films.[24]

The temperature dependent carrier concentration and mobility are shown in Fig.2. Contacts are observed to be not ohmic for Sample A below 160K. We observe no significant carrier freeze out in all the samples down to 80K, indicating degenerate doping. To further confirm this observation, we estimate the Fermi level separation with the conduction band edge at all temperatures, for which accurate band parameters are essential.

We calculate the band dispersion parameters using density functional theory. The effective mass calculations are performed using the Heyd-Scuseria-Ernzerhof (HSE06) screened hybrid functional and projector augmented wave (PAW) pseudopotentials as implemented in the VASP code.[25–27] All calculations adopt a fraction of 32% Hartree-Fock exact exchange, included the Ga 3d states as explicit valence electrons, and adopt a plane wave energy cutoff of 520 eV. The lattice constants for bulk Ga$_2$O$_3$ and the Al-containing alloys with 25% and 50% Al are adopted from literature,[9,28] where Al is only considered to occupy the octahedral Ga site. The 50% Al alloys have full occupation of the octahedral sites, and we investigate two configurations of 25% Al content to assess the role of disorder on the octahedral site for partial occupation. The conduction band effective masses are determined by parabolic and hyperbolic fits along the Γ-X, Γ-Y, Γ-Z, Γ-M, and Γ-N directions in reciprocal space using a $k$-grid spacing of 0.02 Å$^{-1}$ within the primitive unit cells. Specifically, we fit the lowest conduction band in the vicinity of the conduction band minimum (CBM) at Γ using an expression of the form (equation 1)

$$\frac{\hbar^2 k^2}{2m^*} = \epsilon(1 + \alpha\epsilon) \quad (1)$$

where $m^*$ is the electron effective mass at Γ, $\epsilon$ and $k$ are the band energy and momentum, and $\alpha$ is the non-parabolicity parameter, which is set to 0 for parabolic fits.[29] The values along these directions are summarized in Table 2(see supplementary data), where we find highly isotropic effective masses and significant non-parabolicity parameters for binary $Ga_2O_3$ [29] and the Al-containing alloys. We find slight variations in the 25% alloy for the different Al ordering, with effective mass values of 0.31±0.01 (0.22±0.01) and 0.30±0.01 (0.22±0.02) for the direction-averaged $m^*(\alpha)$ values obtained for the two configurations considered. Linear fits of the obtained $m^*$ and $\alpha$ values over the three Al compositions leads equations of the form $m^*(x_{Al}) = 0.28 + 0.11\ x_{Al}$ and $\alpha(x_{Al}) = 0.21 + 0.04\ x_{Al}$ for an alloy of composition $(Al_xGa_{1-x})_2O_3$ for $x_{Al}$ between 0 and 0.5.

To compare more directly with the coherently strained alloys grown epitaxially on (010) substrates, the influence of strain on the effective masses is also considered. The $a$ and $c$ lattice constants are fixed to that of bulk $Ga_2O_3$, with the $b$-axis optimized. The resulting $b$ lattice parameter is decreased by 0.05 Å for the 25% Al alloys, in good agreement with the decrease experimentally observed by Oshima *et al.*[21] We find that the strain did not significantly influence the resulting effective mass values of $m^* = 0.31\pm0.02$ and $\alpha = 0.22\pm0.02$, indicating that the values determined for bulk alloys are also applicable to the coherently strained films.

The fermi level separation from conduction band edge ($E_F - E_C$) is calculated as a function of temperature (Fig. 2(c)) using equation (2),[30] with the DFT calculated parameters, where $\varepsilon_c = (E - E_c)/k_BT$, $\eta = (E_f - E_c)/k_BT$ and $N_c = 2(m_ek_BT/2\pi\ \hbar^2)^{1.5}$,

$$n = \left(\frac{2}{\sqrt{\pi}}\right) N_c \int_0^\infty \frac{(1+2\alpha K_B T\varepsilon_c)*\sqrt{(\varepsilon_c(1+\alpha K_B T\varepsilon_c))}}{1+e^{\varepsilon_c-\eta_c}}\, d\varepsilon_c \qquad (2)$$

Sample A is found to be weakly degenerate and sample B and C are strongly degenerate,

with $E_f$-$E_c$ >> $k_BT$, at all T, for samples B and C. At high carrier concentration parabolic approximation ($\alpha = 0$) overestimates the Fermi level compared to non-parabolic band model. Hence to confirm degenerate doping, it is essential to also include band non-parabolicity for heavily-doped degenerate semiconductors.

The room temperature Hall mobilities (Table 1) of n- β-(Al$_{0.26}$Ga$_{0.74}$)$_2$O$_3$ are found to be in the range of values seen in β-Ga$_2$O$_3$ for similar carrier concentrations.[17,19,31] Temperature dependent Hall measurements (Fig. 2b) indicate that the films suffer from heavy ionized impurity scattering at low temperature, which is expected considering the high density of donor atoms. We hypothesize that the enhanced screening of phonon modes at higher carrier concentration could be the reason for higher mobility measured in sample C compared to sample B.[32] The temperature mobility curve shows a peak ~150K for all the samples and a drop in mobility with increasing temperature, as expected in typical semiconductors.

Next, we study modulation doping in MOVPE-grown β-(Al$_x$Ga$_{1-x}$)$_2$O$_3$/β-Ga$_2$O$_3$ heterostructure. The stack consists of 175 nm UID Ga$_2$O$_3$ layer, 2 nm UID β-(Al$_x$Ga$_{1-x}$)$_2$O$_3$ spacer and 16 nm Si-doped β-(Al$_x$Ga$_{1-x}$)$_2$O$_3$ barrier layer (estimated doping of 1 x 10$^{19}$ cm$^{-3}$) (Fig. 3). The growth is carried out on a Sn-doped substrate in order to make a back contact to the 2DEG channel. The rms surface roughness (5x5 μm$^2$ scan area) of the sample is found to be 1.9 nm (see supplementary data). Aluminum content in the barrier layer is found to be ~26% from HRXRD measurements(Fig.3), this corresponds closely with [TMAl]/[TMAl+TEGa] molar ratio (25%). Film thickness of 18 nm is extracted from the fringe spacing in the HRXRD. Ti/Au (50 nm/50 nm) ohmic contacts are sputtered on the back side of the substrate and Ni/Au (50 nm/50 nm) contacts (300 μm diameter) are deposited as Schottky contacts using e-beam evaporation on the epilayer surface.

CV measurements (100 kHz) are performed on the device to characterize the charge density and depth profile of the electron channel. The measured depth profile shows confinement of carriers at the heterointerface and the carrier concentration rapidly falls off with increasing depth. The peak of the 2DEG layer is close to the estimated barrier thickness from HRXRD measurements, confirming modulation doping. A pinch-off voltage($V_p$) of -1.5V is extracted from the capacitance-voltage profile (see supplementary data), and the total extracted sheet charge from the depth profile is $2.3 \times 10^{12}$ cm$^{-2}$ (Fig. 4a). Low temperature Hall measurements are required to rule out any parallel channel in the alloy barrier, which requires direct formation of ohmic contacts to the 2DEG channel and growth on insulating substrates. Low channel charge necessitates ohmic contact regrowth, which is beyond the scope of this report.[8,14] We also see shift in the 2DEG peak with different β-(Al$_x$Ga$_{1-x}$)$_2$O$_3$ barrier thickness, as expected in modulation doped samples (see supplementary data). The simulated band diagram using a Poisson-Schrödinger solver in included in Fig. 4b, assuming a band offset ($\Delta E_c$) of 0.4eV,[33] a barrier height of 1.9eV, dielectric constant of 10 and a shallow Si donor level in β-(Al$_{0.26}$Ga$_{0.74}$)$_2$O$_3$ layer with a doping density of $1 \times 10^{19}$ cm$^{-3}$, we estimate equilibrium sheet charge $n_s \sim 2.2 \times 10^{12}$ cm$^{-2}$, which is close to the measured value from CV profile.

In summary, we demonstrate growth of heavily-doped β-(Al$_{0.26}$Ga$_{0.74}$)$_2$O$_3$ epilayers using silane as a precursor in MOVPE with room-temperature electron concentration $n \sim 6 \times 10^{18} - 7.3 \times 10^{19}$ cm$^{-3}$ and mobility of 53 – 27 cm$^2$/V.s. We confirm degenerate doping in these epilayers using DFT calculated band parameters including band non-parabolicity. Further, we demonstrate modulation doping in MOVPE grown β-(Al$_{0.26}$Ga$_{0.74}$)$_2$O$_3$/β-Ga$_2$O$_3$ heterojunction and measure a total electron sheet charge of $2.3 \times 10^{12}$ cm$^{-2}$.

**Acknowledgments**

This material is based upon work supported by the Air Force Office of Scientific Research under award number FA9550-18-1-0507 monitored by Dr Ali Sayir. Any opinions, finding, and conclusions or recommendations expressed in this material are those of the author and do not necessarily reflect the views of the United States Air Force. This work was performed in part at the Utah Nanofab sponsored by the College of Engineering and the Office of the Vice President for Research. We also thank Jonathan Ogle, Prof. Luisa Whittaker-Brooks Prof. Michael A. Scarpulla at the University of Utah for providing access to equipment used in this work. We thank Air Force Research Lab for discussions and providing substrates used in this work. The theoretical work was performed under the auspices of the U.S. DOE by Lawrence Livermore National Laboratory under contract DE-AC52-07NA27344, and supported by the Critical Materials Institute, an Energy Innovation Hub funded by the U.S. DOE, Office of Energy Efficiency and Renewable Energy, Advanced Manufacturing Office.

**Figure Captions**

**Fig. 1.** HRXRD 2θ - ω measurements for samples A, B and C; inset showing epitaxial structure of β-(Al$_x$Ga$_{1-x}$)$_2$O$_3$ thin films grown by MOVPE.

**Fig. 2.** a) Low temperature Hall carrier concentration of samples A, B and C, b) Temperature dependent Hall mobility of samples A, B and C between 80 – 340 K, c) calculated Fermi level separation from conduction band edge, considering α = 0.22 (non-parabolic bands) and α = 0 (parabolic bands) for samples A, B and C

**Fig. 3.** HRXRD scan of modulation-doped β-(Al$_{0.26}$Ga$_{0.74}$)$_2$O$_3$/ β-Ga$_2$O$_3$ heterojunction with inset showing the epitaxial structure.

**Fig. 4.** a) Carrier concentration vs depth profile extracted from CV measurement of β-(Al$_{0.26}$Ga$_{0.74}$)$_2$O$_3$/ β-Ga$_2$O$_3$ heterojunction, b) Calculated equilibrium band diagram for modulation-doped β-(Al$_{0.26}$Ga$_{0.74}$)$_2$O$_3$/β-Ga$_2$O$_3$ heterojunction.

**Table I.** Electrical characterization of doped $(Al_{0.26}Ga_{0.74})_2O_3$ thin films

| Sample | Growth time (minutes) | Silane Molar Flow(nmol/min) | Thickness of $\beta\text{-}(Al_{0.26}Ga_{0.74})_2O_3$ (nm) | RT Carrier Concentration (cm$^{-3}$) | RT Hall Mobility (cm$^2$/V.s) | RT Resistivity ($\Omega$-cm) |
|--------|-----------------------|------------------------------|-----------------------------------------------------------|--------------------------------------|-------------------------------|------------------------------|
| A | 9 | 1.94 | 30 | $6 \times 10^{18}$ | 53 | 0.019 |
| B | 12 | 15.18 | 40 | $3.5 \times 10^{19}$ | 27 | 0.0066 |
| C | 12 | 50.54 | 40 | $7.3 \times 10^{19}$ | 32 | 0.0026 |

Template for APEX (Jan. 2014)

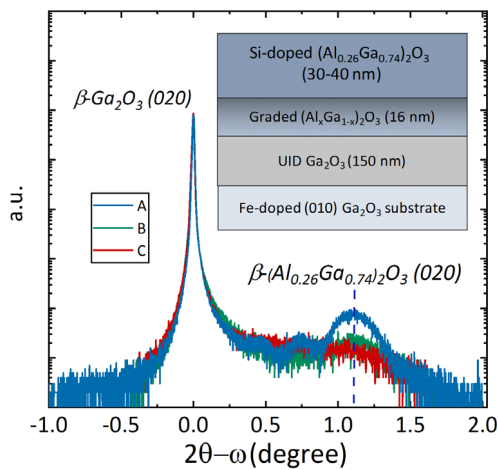

**Fig. 1.** HRXRD 2θ-ω measurements for samples A, B and C; inset showing epitaxial structure of $(Al_xGa_{1-x})_2O_3$ thin films grown by MOVPE.

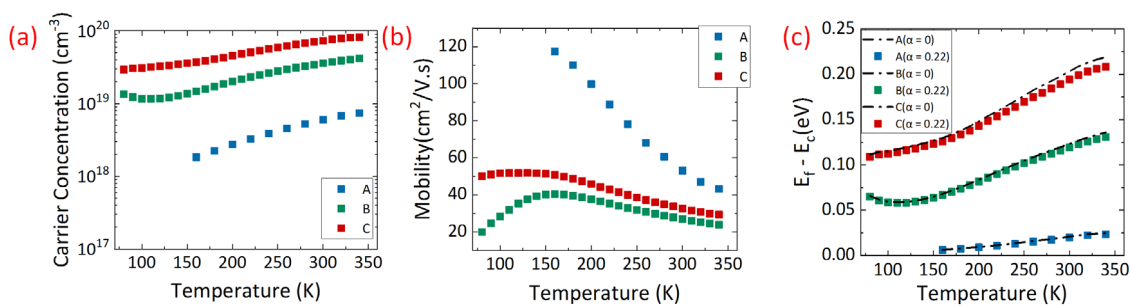

**Fig. 2.** a) Low temperature Hall carrier concentration of samples A, B and C, b) Temperature dependent Hall mobility of samples A, B and C between 80 – 340 K, c) Calculated Fermi level separation from conduction band edge, considering α = 0.22 (non-parabolic bands) and α = 0 (parabolic bands) for samples A, B and C

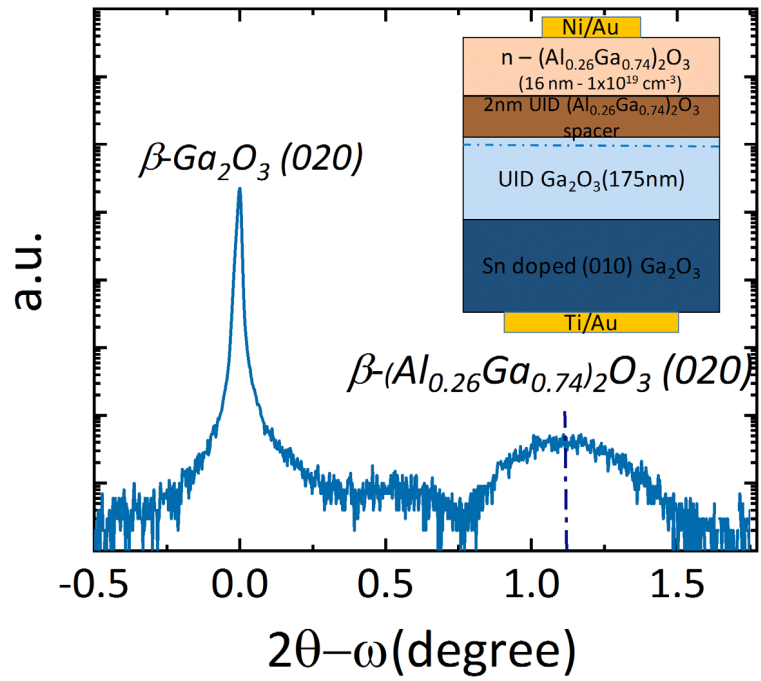

**Fig. 3.** HRXRD scan of modulation-doped β-(Al$_{0.26}$Ga$_{0.74}$)$_2$O$_3$/β-Ga$_2$O$_3$ heterojunction with inset showing the epitaxial structure.

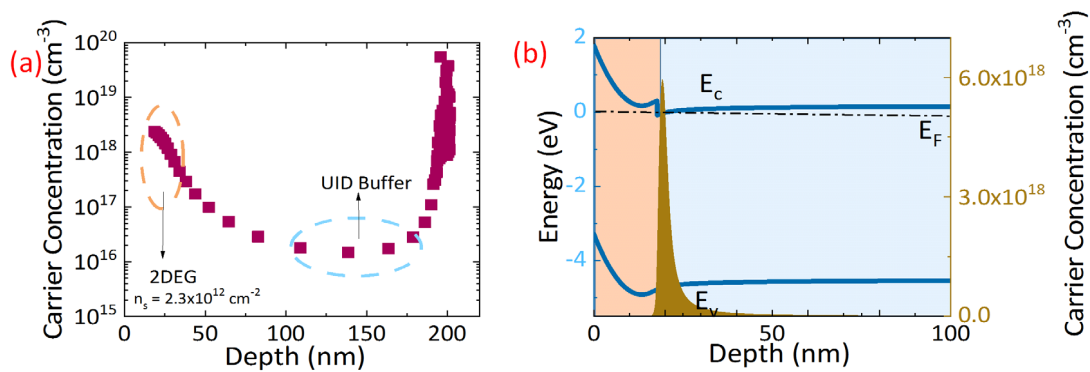

**Fig. 4.** a) Carrier concentration vs depth profile extracted from CV measurement of β-(Al$_{0.26}$Ga$_{0.74}$)$_2$O$_3$/ β-Ga$_2$O$_3$ heterojunction, b) Calculated equilibrium band diagram for modulation-doped β-(Al$_{0.26}$Ga$_{0.74}$)$_2$O$_3$/β-Ga$_2$O$_3$ heterojunction.